\documentclass[unnumsec,webpdf,contemporary,large]{LaTeX_Suppl_Files/oup-authoring-template}%

\onecolumn 

\graphicspath{{Fig/}}

\usepackage{graphicx}
\usepackage{subcaption}

\usepackage{newunicodechar}
\newunicodechar{​}{}  

\makeatletter
\renewcommand\paragraph{\@startsection{paragraph}{4}{\z@}%
	{-3.25ex\@plus -1ex \@minus -.2ex}%
	{1.5ex \@plus .2ex}%
	{\normalfont\normalsize\bfseries}}
\makeatother

\theoremstyle{thmstyleone}%
\theoremstyle{thmstyletwo}%
\theoremstyle{thmstylethree}%

\usepackage{pdflscape}
\usepackage{booktabs}
\usepackage{adjustbox} 

\usepackage{setspace}
\usepackage{lineno}

\begin{document}
	
\journaltitle{XXXX XX XXXX}
\DOI{DOI HERE}
\copyrightyear{2023}
\pubyear{2019}
\access{Advance Access Publication Date: Day Month Year}
\appnotes{Review}

\firstpage{1}

\title[Human-AI Collaborative Genome Annotation]{A Conceptual Framework for Human-AI Collaborative Genome Annotation}

\author[1,$\ast$]{Xiaomei Li\ORCID{0000-0002-8870-3186}}
\author[2]{Alex Whan}
\author[3]{Meredith McNeil}
\author[4]{David Starns}
\author[5]{Jessica Irons}
\author[1]{Samuel C. Andrew}
\author[6,7, $\ast$]{Rad Suchecki}

\authormark{Li et al.}

\address[1]{\orgdiv{AGRICULTURE \& FOOD}, \orgname{CSIRO}, \orgaddress{\street{26 Pembroke Rd}, \postcode{2122}, \state{NSW}, \country{Australia}}}
\address[2]{\orgdiv{AGRICULTURE \& FOOD}, \orgname{CSIRO}, \orgaddress{\street{2-40 Clunies Ross Street, Acton}, \postcode{2601}, \state{ACT}, \country{Australia}}}
\address[3]{\orgdiv{AGRICULTURE \& FOOD}, \orgname{CSIRO}, \orgaddress{\street{306 Carmody Road, St Lucia}, \postcode{4067}, \state{QLD}, \country{Australia}}}
\address[4]{\orgdiv{School of Molecular and Cellular Biology}, \orgname{University of Leeds}, \orgaddress{\street{Woodhouse Lane}, \postcode{LS2 9JT}, \state{Leeds}, \country{United Kingdom}}}
\address[5]{\orgdiv{Data 61}, \orgname{CSIRO}, \orgaddress{\street{13 Garden Street, Eveleigh}, \postcode{2015}, \state{NSW}, \country{Australia}}}
\address[6]{\orgdiv{AGRICULTURE \& FOOD}, \orgname{CSIRO}, \orgaddress{\street{Waite Road, Urrbrae}, \postcode{5064}, \state{SA}, \country{Australia}}}
\address[7]{\orgdiv{}, \orgname{Alkahest Inc.}, \orgaddress{\street{125 Shoreway Rd D, San Carlos}, \postcode{94070}, \state{CA}, \country{United States}}}

\corresp[$\ast$]{Corresponding author. \href{email:Maisie.Li@csiro.au}{Maisie.Li@csiro.au}, \href{email:rsuchecki@alkahest.com}{rsuchecki@alkahest.com}.}

\received{Date}{0}{Year}
\revised{Date}{0}{Year}
\accepted{Date}{0}{Year}



\abstract{Genome annotation is essential for understanding the functional elements within genomes.  
While automated methods are indispensable for processing large-scale genomic data, they often face challenges in accurately predicting gene structures and functions. 
Consequently, manual curation by domain experts remains crucial for validating and refining these predictions. 
These combined outcomes from automated tools and manual curation highlight the importance of integrating human expertise with AI capabilities to improve both the accuracy and efficiency of genome annotation. 
However, the manual curation process is inherently labor-intensive and time-consuming, making it difficult to scale for large datasets. 
To address these challenges, we propose a conceptual framework, Human-AI Collaborative Genome Annotation (HAICoGA), which leverages the synergistic partnership between humans and artificial intelligence to enhance human capabilities and accelerate the genome annotation process. 
Additionally, we explore the potential of integrating Large Language Models (LLMs) into this framework to support and augment specific tasks. Finally, we discuss emerging challenges and outline open research questions to guide further exploration in this area.
}


\maketitle

\section{Introduction}\label{sec-intro}

Genome annotation (GA) is the process of identifying and interpreting the functional elements encoded within a genome. It is a critical step in understanding an organism's biology, enabling researchers to connect genetic information to phenotypes, understand disease mechanisms, and uncover evolutionary relationships.
GA is heavily relies on
automated methods, including Machine Learning (ML)
\citep{eraslan2019deep, zou2019primer, mahood2020machine} and other
computational methods such as rule-based and heuristic methods
\citep{altschul1990basic, stanke2006augustus}. However, automated
methods are generally hampered by the relative scarcity of reliable
labeled data and the complexity of biological systems. In fact, gene annotations, particularly
functional annotations, are mostly transferred from one species to
another in an automated manner, relying mainly on the similarity of
underlying nucleotide sequences, or the corresponding protein sequences.

Manual curation is widely recognized as essential for improving the reliability and accuracy of genome annotation \citep{yandell2012beginner, madupu2010meeting, tatusova2016ncbi}. It involves human experts reviewing and refining annotations, particularly by addressing ambiguities or gaps that automated pipelines may overlook. For instance, curators enhance the functional annotations of genes by incorporating new insights from scientific literature that detail experimental results related to gene function. Additionally, manual curation enables precise gene structure annotation by reviewing evidence from multiple sources, such as omics datasets and experimental outcomes, to accurately define gene boundaries \citep{zerbino2020progress}.

Despite their value, these evidence sources are often scattered across multiple platforms or embedded within vast datasets, making manual curation a time-consuming and labor-intensive process. Consequently, current GA practices rely heavily on automated annotation which is not always followed by manual curation \citep{biology9090295}.

Manual curation has mostly been conducted in cases where teams of annotators collaborate to create accurate and up-to-date annotations for high-priority species or specific gene sets.
Correspondingly, computational tools and platforms have been developed to support collaboration among annotators. These tools include algorithms that identify problematic annotations and prioritize them for review \citep{tello2019double}, as well as platforms that facilitate seamless communication, data sharing, and coordination among annotators, regardless of their geographic location \citep{Lee2013, haas2005complete}.
However, these tools operate independently and have not addressed the issue of dispersed data sources across various platforms. Additionally, there is a lack of dynamic interaction between the tools and users, i.e., the tools typically run automatically without a user in the loop. 

As highlighted by Mac \emph{et al.} \citep{mac2014putting}, the disconnect between AI tools and their users can potentially impact the effective utilization of AI. They suggest integrating scientists into the loop and combining their expertise with interactive machine learning to accelerate scientific progress.
There is a growing body of research on how humans and AI collaborate to drive advancements in fields such as medicine \citep{gao2024empowering, van2021biological, tschandl2020human} and chemistry \citep{m2024augmenting}, particularly through the adoption of LLMs. Furthermore, the emerging concept of human-AI \emph{collaborative intelligence} focus on the combination of humans and AIs working together to solve
problems, leveraging the strengths of both parties and enhancing each
other's capabilities \citep{wilson2018collaborative, schleiger2023collaborative}.
Although still in its early stages, an increasing number of studies demonstrate that human-AI collaboration can lead to superior performance in accomplishing complex tasks \citep{goldberg2019robots, wilson2018collaborative, huang2022framework}.

Inspired by these work, we propose a conceptual framework named Human-AI Collaborative Genome Annotation (HAICoGA), in which humans and AI systems not only work interdependently but also actively collaborate over a sustained period.
This iterative collaboration allows domain experts to refine AI outputs, correct errors, and ensure that annotations align with biological context, thereby improving the accuracy of GA.
Moreover, this collaboration may enhance AI itself, as the iterative feedback loop enables an AI system to learn from expert corrections, refine its model, and better generalize to new data.
Finally, AI systems are proposed to integrate the necessary tools and resources for GA into a unified platform, streamlining the annotation process. By consolidating multiple functionalities, this platform reduces fragmentation and enhances the efficiency of the GA workflow.

The remainder of this paper is organized as follows. Section 2 provides the necessary background and reviews related work relevant to this study. Section 3 introduces the conceptual framework of HAICoGA, identifying key components, critical capabilities and mechanisms required to establish an effective and sustainable human-AI collaborative relationship. In Section 4, we explore current applications of LLM-based AI agents in the biological and biomedical domains and present a vision for the HAICoGA workflow. Section 5 outlines key future research directions to further realize HAICoGA. We hope this work contributes to the development of human-AI collaborative workflows for GA in the future.

\section{Background and related work}\label{genome-annotation-workflow}

\subsection{Genome annotation}\label{genome-annotation}

GA can be interpreted as multi-dimensional, spanning from the nucleotide level to the
biological system level \citep{reed2006towards}. Genomic elements of
interest include, but are not limited to, single nucleotide polymorphisms
(SNPs), coding genes, non-coding genes, regulatory elements and other
non-coding regions. Structural annotation primarily focuses on
delineating the physical regions of genomic elements. While the structural annotation offers initial
clues, a definitive understanding of functions still requires in-depth
analysis. 

GA encompasses a broad range of tasks that are now primarily accomplished through 
various computational approaches utilizing diverse data types. We
provide a rough chronology of the emergence and prominence of different 
automated methods in Supplementary Note 1. These
automated methods can be integrated into highly complex pipelines to
perform multiple steps in automated GA. Although automated approaches
dominate GA, they still faces serious limitations and challenges
(Supplementary Note 1).

Due to the limitations of existing computational tools, automated GA often results in a high rate of erroneous annotations. Such errors include, but are not limited to, incorrect gene boundaries, inaccurate exon-intron structures, misidentification of pseudogene regions as genes, and improper functional assignments.
These tools struggle to accurately annotate genes, particularly in non-model species \citep{de2022roadmap}. In many cases, genes in non-model organisms are either assigned annotations borrowed from homologs or labeled with generic terms such as ``hypothetical gene" or ``expressed protein", where functional descriptions lack sufficient detail to capture the gene's biological role.

\subsection{Manual curation}\label{manual-curation}

Manual curation has primarily been done in model species to continuously
improve the accuracy and coverage of their genome annotations. For
example, projects such as HAVANA for the human genome, TIGR
for~\emph{Arabidopsis thaliana}, and ITAG for~\emph{Solanum
lycopersicum} produce high-quality annotations manually curated by
specialized experts. Manual curation is not limited to collaborative efforts or decentralized networks (detailed manual curation models are provided in Supplementary Note 2); it also plays a crucial role in individual research. Researchers may engage in manual curation before formulating hypotheses for their studies or when interpreting their results.

Manual curation is an ongoing process that requires the continuous
repetition of five general steps
\citep{cheng2017araport11, lamesch2012arabidopsis} (see Supplementary Note
2). This process is time- and labour-intensive, but it can be
made more efficient with the assistance of software tools. 

For example, Apollo has been widely used in the GA community for quite some time and is continually updated \citep{Lewis2002, Lee2013, dunn2019apollo}. The current web version allows users to perform real-time collaborative annotation.
A fast and scalable genome browser, JBrowse \citep{buels2016jbrowse}, is integrated into Apollo for visualizing genomes and diverse evidence tracks, such as RNA-seq data, short peptides, and gene models from different gene prediction tools. Additionally, a sequence alignment tool, BLAT \citep{kent2002blat}, is incorporated into Apollo to help users quickly identify the loci of their specific genes or sequence fragments of interest.
Users can easily edit gene models within Apollo, including adding gene models or exons, adjusting exon boundaries, merging and splitting genes, and deleting gene models or exons. While Apollo enables curators to add gene functional annotations, they must rely on external tools, such as text mining tools, to find gene functions and supporting evidence, as Apollo does not provide these functionalities natively.

Text-mining tools have been widely used to efficiently extract information about valuable biological entities, such as genes, proteins, and detailed aspects of gene functions. For example, PubTator Center provides a web service that annotates multiple biological entities in full-text articles \citep{wei2019pubtator}.
Manually curating functional annotations from the literature remains a challenging task, as highlighted by the complexities addressed in competitions such as BioCreative IV \citep{mao2014overview}. Despite advances in automation, this process still requires significant expertise and manual effort from annotators, who need to carefully validate the information provided by text-mining tools \citep{drabkin2012manual}.

More details on additional tools can be found in Supplementary Note 2. These tools offer diverse functionalities to assist manual curation, including platforms for information collection, interactive visualization tools for structural annotation, and tools for curating functional annotations from scientific literature.
However, these tools are distributed across different platforms. Additionally, automated GA tools operate independently from these manual curation tools. As a result, humans need to spend a significant amount of time running and navigating multiple tools, as well as transferring data between them.
To accelerate the GA process, an integrated system is needed, which connects all necessary automated and manual GA tools and enables seamless collaboration between humans and AI tools.

\subsection{Human-AI collaboration}
AI systems can play different roles in human-AI teaming, including automation, augmentation, and collaboration \citep{baruwal2024towards}. In automation, AI independently performs tasks without human intervention. In augmentation, AI enhances human experts' abilities in their tasks. Collaboration refers to humans and AI working together in a coordinated effort toward shared goals, enabling better outcomes than either could achieve alone. Similarly, Gao \emph{et al.} \citep{gao2024empowering} classify AI systems into four intelligence levels based on their capabilities in hypothesis generation, experimentation, and reasoning. Level 0 consists of AI models used by humans as automated tools. Level 1 includes AI assistants that execute tasks specified by scientists. Level 2 consists of AI collaborators that work alongside scientists to refine hypotheses and utilize a broader array of tools for experimentation and discovery. Level 3 represents AI scientists, which exhibit the highest level of intelligence.

In the current GA workflow, automated GA tools can be categorized as Level 0 AI models. Recently, LLMs have been used to develop various AI assistants (Level 1) in the biological and biomedical domains, which could potentially be adapted for GA.
However, a significant gap remains due to the lack of human-AI collaboration in GA (Level 2). To address this, we propose a conceptual framework that integrates the automation, augmentation, and collaboration roles of AI systems, allowing humans and AI models to work interdependently or collaboratively.

There are related research areas, such as human-in-the-loop machine learning (ML) \citep{mosqueira2022human}, human-centered AI \citep{riedl2019human}, and human-AI teaming \citep{bansal2019updates}. Human-in-the-loop ML focuses on scenarios where AI alone may struggle to achieve optimal results, requiring humans to provide guidance, feedback, or decision-making inputs to the AI system.
Human-centered AI emphasizes that humans retain primary decision-making authority, utilizing AI as an augmentative tool to enhance human capabilities without superseding the human role. 
Human-AI teaming refers to an organizational setup in which AI systems are increasingly considered team members rather than mere support tools for humans. 
There are overlaps among these research areas, and these concepts may be used interchangeably in the literature. 

The emergence of human-AI collaboration (HAIC) frameworks and taxonomies is contributing to a deeper understanding of the fundamentals of human-AI collaboration \citep{dubey2020haco, dellermann2021future}. Several HAIC frameworks have been introduced across various domains, including manufacturing, healthcare, finance, and education.
For example, Hartikainen \emph{et al.} \citep{hartikainen2024human} present a framework that provides an initial basis for designing human-AI collaborative systems in smart manufacturing. Similarly, Viros i Martin \emph{et al.} \citep{viros2021framework} propose a framework to describe and explain human-AI collaborative design, with a focus on Design Space Exploration. Additionally, Dubey \emph{et al.} \citep{dubey2020haco} present a framework for developing successful human-AI teaming and demonstrate its use in contact centers.
Despite its broad potential, applying HAIC to GA presents unique challenges. Therefore, we extend recent studies in HAIC to GA and propose future research directions for developing HAICoGA workflows.

\section{A conceptual framework of human-AI collaborative genome annotation}
\begin{figure}[]
\centering
\includegraphics[width=0.9\textwidth]{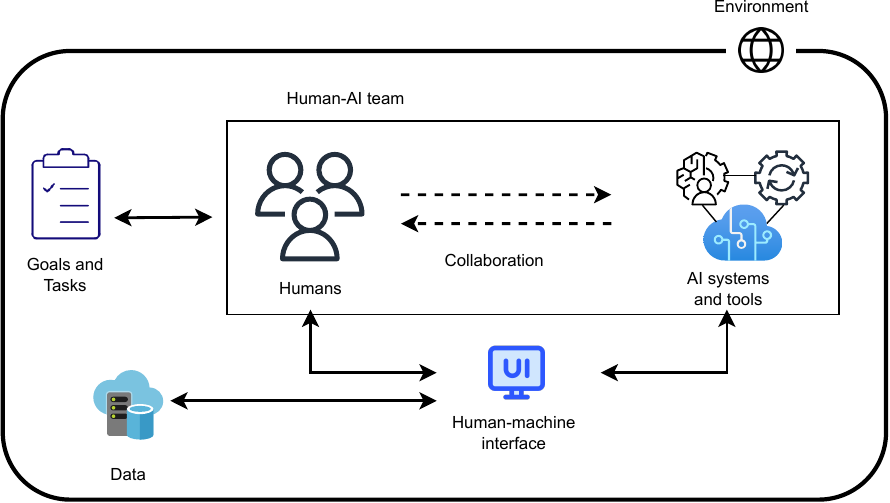}
\caption{Key elements in human-AI collaborative genome annotation. Humans and AI work together as a team to perceive the environment in which they operate. To achieve the shared high-level goal, they decompose the task into sub-tasks and objectives. Through the human-machine interface, humans and AI utilize the available data to carry out tasks and transition to a new state within the environment. This state may lead to updates in the list of tasks and goals or the addition of new data until the final goal is achieved. The collaboration between humans and AI is dynamic, allowing them to perform individual tasks independently while collaborating on shared tasks when necessary.}
\label{HAICoGA}
\end{figure}

First, we present the most relevant elements considered in HAICoGA by extending the concepts and taxonomy from general theories in HAIC \citep{dubey2020haco, dellermann2021future, hartikainen2024human}. As illustrated in Figure \ref{HAICoGA}, HAICoGA consists of seven key elements: humans, AI systems and tools, data, goals and tasks, human-machine interface, environment, and collaboration.

\subsection{Key elements}
\label{framework-for-human-ai-collaborative-genome-annotation}

\paragraph{Humans}
Humans refers to individuals involved in GA, particularly those engaged in manual curation. This group may include biological researchers, experimental scientists, biocurators, and trained students. These individuals may be dispersed across different geographic locations but collaborate toward a shared task.

Humans have diverse knowledge,
skills and expertise that they bring to the decision-making process
\citep{cooper2017cognitive}. Research suggests experts rely heavily on
their previous experience on the task, as well as their deep and often
tacit knowledge of the domain \citep{klein2008naturalistic}.
Decision-makers often employ mental shortcuts and heuristics that are
efficient but can be prone to cognitive bias
\citep{kahneman1982judgment, simon1977scientific}.

While humans lack the
information processing capacity of AI, they are sensitive to changing
task goals and context, enabling flexible and adaptive decision-making
\citep{klein2008naturalistic}. 
Due to these unique human capabilities, their role cannot be completely replaced by AI \citep{paris2024secret}.

\paragraph{AI systems and tools}
The AI element consists of a collection of systems and tools designed to perform or assist in GA. By using the term ``AI," we assume that computational tools for GA can be integrated into an AI system to enhance collaboration with humans in performing GA tasks. The AI element should comprise an ecosystem of AI tools with diverse designs and functionalities.

Specifically, automation AI tools streamline the multi-step process of identifying and categorizing genes and other functional elements within a genome, including GA pipelines and automation methods \citep{mungall2002integrated, stanke2006augustus}. Augmentation AI tools assist humans in performing their tasks more efficiently, enabling them to work better and faster than they could alone, for example, Apollo \citep{Lee2013}. Collaboration AI tools facilitate bi-directional human-AI interactions, seamlessly integrating AI's computational strengths with human intuition and expertise, thereby potentially improves
the overall effectiveness of the GA process. 
These AI tools can also be interconnected with one another, forming an integrated system for enhanced functionality.

\paragraph{Data}
Sequence data play an important role in GA, including genome assemblies, expressed sequence tags (ESTs), complementary DNAs (cDNAs), RNA-seq data, and protein sequences \citep{biology9090295}. Biological databases, such as UniProt \citep{uniprot2023uniprot} and Gene Ontology (GO) \citep{gene2021gene}, aggregate knowledge from experiments and studies, providing labeled data for GA. Scientific publications serve as contextual evidence for gene annotations.

Knowledge graphs (KGs) are also valuable data sources for GA. For example, GenomicKB \citep{feng2023genomickb} can be used to identify genes and genetic variants related to human diseases, while KnetMiner \citep{hassaniknetminer} supports the identification of critical genes associated with complex traits and diseases across plant species.

Beyond data directly used for GA, additional data provide information about the context and environment in which human-AI collaboration occurs. For instance, data on how users interact with the AI system, including the queries they input, the options they select, and the feedback they provide. These data help refine AI algorithms to better suit user needs.

\paragraph{Goals and Tasks}
HAICoGA is built upon the collaborative partnership between humans and AI to accomplish shared goals. These goals can be both individual and collective \cite{fragiadakis2024evaluating}. AI's individual goals may include identifying repetitive elements, genes, and their functions. Humans review the AI's predictions and generate hypotheses. Assigning biological functions to genes serves as a shared overall goal between humans and AI.

The overall goal of a task can be achieved through subtasks and structured plans. In hierarchical task analysis, a task is broken down into subtasks until a stop criterion is reached, often when the subtask consists of only a single operation \citep{bligaard2014ccpe}. A single operation, such as gene prediction, can be accomplished by an individual AI, while other single operations may be performed by humans. These subtasks can be assigned to humans and AI based on their capabilities.

In some use cases, tasks are allocated dynamically, and responsibilities shift based on real-time needs \cite{fragiadakis2024evaluating}. Plans organize a sequence of tasks either sequentially or hierarchically and help allocate individual humans or AI to specific tasks \citep{bligaard2014ccpe}.

\paragraph{Human-machine interface}
The human-machine interface facilitates interactions between humans and AI, a fundamental basis of collaborative AI. One key factor in enhancing user experience is the provision of user-friendly interfaces. Graphical user interfaces (GUIs) abstract away the complexities of command-line programming, allowing users without coding experience to leverage advanced AI models. GUIs also provide a visual and interactive means for humans to engage with AI systems in GA.

For example, GUIs are used to visualize feature locations, display evidence alignments, and present other relevant information in genome browsers such as GBrowse \citep{stein2013using}, UCSC Genome Browser \citep{kent2002human}, NCBI Genome Data Viewer (GDV) \citep{rangwala2021accessing}, and JBrowse \citep{buels2016jbrowse}. Through these genome browsers, users can compare predicted gene structures with evidence tracks to detect errors in genome annotations \citep{Lee2013}. GUIs like Apollo not only visualize genomic data but also allow users to directly edit gene annotations within the system.

Recent advancements in LLMs have significantly contributed to the growing interest in Conversational User Interfaces (CUIs). CUIs allow users to interact with computers using natural language, making it easier to access information and perform tasks without needing to learn complex commands or navigate intricate menus. For certain tasks, CUIs enhance human-AI collaboration by enabling effective and intuitive interactions between humans and AI \citep{feng2024large}.

\paragraph{Environment}
The environment encompasses the physical, organizational, and psychosocial spaces where humans and AI agents work together \citep{bligaard2014ccpe}. It involves multiple participants and complex scenarios, categorized into the digital, task, and team environments.

The digital environment includes conditions and factors such as software platforms, interface design, and the availability of data for both humans and AI \citep{madni2018architectural}. The task environment pertains to the tasks that need to be completed, the constraints and limitations involved, and the desired outcomes \citep{NAS2022HumanAITeaming}. The team environment refers to the dynamics and structures within a group of individuals (including both human and AI) working together \citep{salas2015understanding}. It is characterized by the roles and relationships established among team members, communication patterns, and the level of cooperation and collaboration required to achieve common goals.

The environment significantly influences how humans perceive and interact with AI, as well as AI's ability to understand and respond to human input. Understanding the environment is crucial for designing effective human-AI interactions.

\paragraph{Collaboration}
Collaboration is a central element of HAICoGA. Successful collaboration relies heavily on effective communication and feedback mechanisms between humans and AI. Research on human-machine collaboration is supported when both the human and machine have a common goal and they share an understanding of the task and the team make-up (including each team-member's capabilities and limitations, e.g. \citep{lyons2021human, o2022human}).

The aspiration for HAICoGA is for humans and AI to work together, leveraging their unique strengths and expertise to solve GA problems more efficiently and effectively. Their complementary strengths create opportunities to exploit synergistic potential, enabling them to achieve goals that neither could accomplish individually \citep{schleiger2023collaborative}. However, the mechanisms by which humans and AI operate differ, posing integration challenges.

\subsection{Capabilities of Humans and AI}
It is essential to understand the different ways in which humans and machines operate, and the unique competencies each bring to GA (Figure \ref{HAI}, detailed in Supplementary Notes 3-4.). 
Human cognitive mechanisms, such as learning, reasoning, situational awareness, decision-making, delegation, and trust calibration, that play a pivotal role in collaborative systems (Supplementary Note 3). 
Human strengths include abstract thinking, contextual understanding, and dynamic problem-solving, which AI currently lacks but can learn from or emulate to a limited extent.

\begin{figure}[]
\centering
\includegraphics[width=0.9\textwidth]{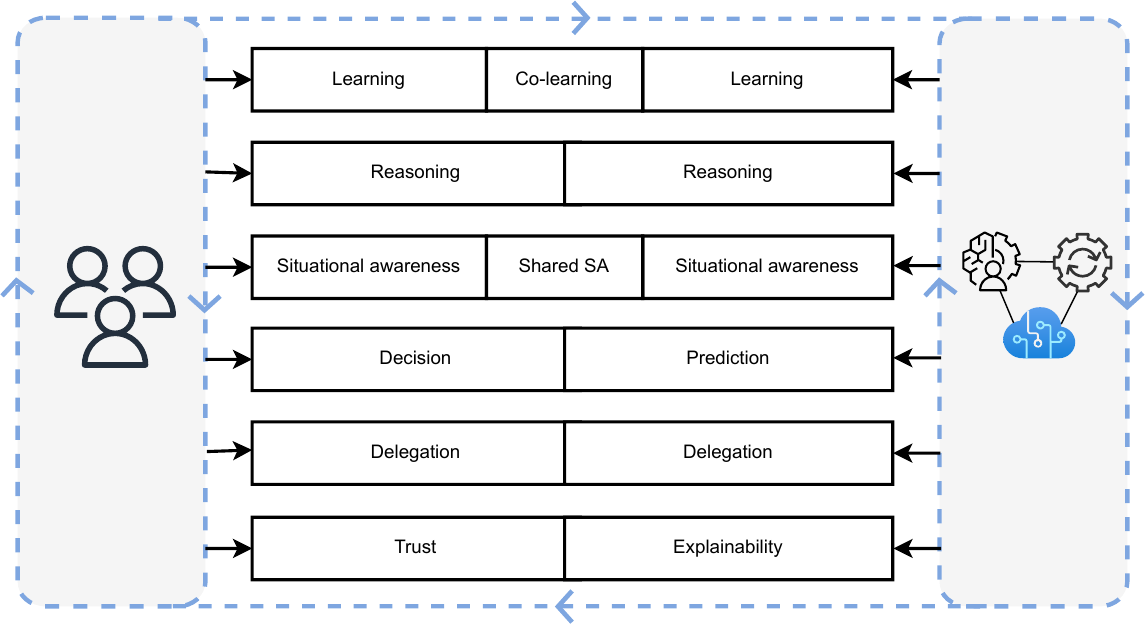}
\caption{Key competencies for fostering effective for human-AI collaboration. SA: situational awareness.}
\label{HAI}
\end{figure}

In the context of GA, AI brings unique competencies that complement human strengths, including rapid learning from data and predictive modeling. Learning strategies such as reinforcement learning, continual learning, and active learning enable AI to collaborate with humans and dynamically adapt to annotation challenges.
For example, through active learning, AI can identify genomic regions where annotation uncertainty is high, directing human curators' attention to those areas of the genome. Therefore, AI not only provides predictions about gene structures and functions but also offers insights into areas requiring further investigation. Crucially, AI can delegate tasks beyond their confidence threshold, such as short genes or complex duplications, to human curators.

AI reasoning thrives on logical processing and scalability. However, advances in causal reasoning and explainable AI are enabling AI to better emulate human-like intuition and contextual understanding, making their outputs more interpretable and aligned with human decision-making processes. AI systems should be designed to maintain situational awareness, ensuring alignment with human collaborators' goals and the ability to adapt to real-time changes in priorities or environmental conditions.
A detailed exploration of these competencies is provided in Supplementary Note 4. 

\subsection{Feedback loop mechanism}
Humans and AI interact dynamically through bi-directional interactions, which are structured as feedback loops (dashed lines in Figure \ref{HAI}). These feedback loops continuously enhance the capabilities of both humans and AI systems.
For example, human-AI interactions facilitate efficient learning through various learning paradigms, such as learning from human demonstrations, learning from human interventions, and learning from human evaluations \cite{waytowich2018cycle}. Specifically, humans provide feedback on AI outputs by correcting errors, validating results, or suggesting alternative approaches. AI then analyzes this feedback and adjusts its algorithms or models accordingly, aiming to produce more accurate, reliable, and contextually appropriate results over time.
This iterative process enables ongoing improvement and adaptation, allowing AI to better align with human expectations, requirements, and preferences, as demonstrated by InstructGPT \citep{ouyang2022training}.

\section{AI agents bring opportunities to realize the HAICoGA framework}
There is an emerging trend of using LLM-backed AI assistants for research. These AI assistants can ``understand" human language and respond to humans in a manner similar to human communication.
In this section, we summarize recent research on the use of AI assistants in biological and biomedical domains and then highlight the opportunities in achieving AI collaboration.

\subsection{LLM-based AI assistants in biological and biomedical domains}
In Table 1, an agent refers to an AI system capable of interacting with humans or other agents and using tools to accomplish its tasks. These agents potentially possess cognitive abilities such as perception, reasoning, learning, and planning, enabling them to collaborate with humans. While these abilities may not be as sophisticated as human cognition, they represent a significant step toward the development of more collaborative and intelligent AI systems. We categorize papers based on the number of agents involved.

Some works in Table 1, such as ChatNT, DRAGON-AI, and GeneGPT, are LLM-based models that take human language as input and generate direct answers without involving a cognitive agent. 
ChatNT is a multimodal AI system that integrates DNA, RNA, and protein sequences with neural language processing to solve various genomics tasks, including binary and multi-label classification tasks related to histone and chromatin features, promoter and enhancer regulatory elements, and splicing sites. 
It accepts one or multiple genomic sequences and a prompt as input and returns a textual response. 
ChatNT employs self-supervised learning during the pre-training of DNA encoders and supervised learning during the fine-tuning phase for specific genomic tasks. Many similar works are discussed in the review paper by Zhang \emph{et. al} \citep{zhang2024scientific}.
DRAGON-AI is a method that automatically generates ontology objects based on partial information from a user. All ontology terms and additional contextual information are translated into vector embeddings and indexed. Relevant contextual information is retrieved using a retrieval-augmented generation (RAG) approach and added to construct a prompt, which is then passed as input to an LLM. The LLM completes the term object accordingly.
GeneGPT uses few-shot learning to teach LLMs how to generate web APIs for the National Center for Biotechnology Information (NCBI) databases and answer biological questions based on retrieved information from NCBI databases.

Phenomics Assistant builds an agent to call external tools based on user queries. It helps non-expert users query and interact with complex data from the Monarch Knowledge Graph. VarChat supports genetic professionals by providing concise summaries of scientific literature related to specific genomic variants. It interacts with external databases and utilizes user inputs to guide its querying and summarization processes.
Both Phenomics Assistant and VarChat use a single-agent framework to provide a CUI that interacts with users and has the ability to use tools to solve user questions based on dynamic situations. The conversation history in the chat allows the agent to be aware of the user's state within tasks and incorporate feedback from external tools. Both systems also provide sources for the information in their responses, improving transparency in their processes.

Two-agent systems, ChatGSE, BioDiscoveryAgent, and GeneAgent, consist of a primary agent that interprets the user's query and selects appropriate tools to solve the problem, and a secondary agent that critically evaluates the results or verifies the factual accuracy of the output. The tools either retrieve and process information from various APIs to access online databases or scientific literature. The retrieved information is treated as a source to determine whether the answer is factually accurate compared to the original data. 
Keeping track of intermediate results from tools and the verification process enhances the agent's awareness of the current task status, potentially allowing it to adjust its actions accordingly in the next round of experiments. ChatGSE employs chain-of-thought reasoning to improve its problem-solving success. BioDiscoveryAgent follows the Reflection-Research Plan-Solution framework to enhance its reasoning capabilities. Both ChatGSE and BioDiscoveryAgent also incorporate self-verification mechanisms.
These two agents operate in a sequential manner. All three systems provide some level of explainability by delivering context-rich answers that include references to data sources, literature, or verification reports. GeneAgent, which applies an AI agent for gene set enrichment analysis, focuses on autonomous interactions with domain-specific databases, followed by subsequent LLM verification.

Multi-agent systems are becoming increasingly popular for solving complex problems. These systems integrate multiple AI agents to automate and enhance critical workflows, significantly improving the speed and efficacy of tasks such as gene enrichment analysis, literature searches, and software pipeline executions. For instance, the BRAD system employs a hierarchical structure of agents to manage tasks like literature retrieval and enrichment analysis automation. These agents use a combination of in-context learning and a specialized planner to distribute and organize tasks efficiently. Another example is the BKGAgent, which focuses on knowledge graph checking by querying knowledge graphs, verifying the accuracy of information through external literature or databases, and identifying factual discrepancies. The system's ability to dynamically query and cross-reference structured knowledge graphs and unstructured scientific texts illustrates the integration of RAG, ensuring relevance and context awareness throughout the information processing stages.

Similarly, GenoAgent and TAIS are tailored for analyzing gene expression data from sources like the Gene Expression Omnibus (GEO) and The Cancer Genome Atlas (TCGA). These systems leverage instruction learning and structured prompting to adapt their actions based on feedback and intermediate results, facilitating an iterative correction process that ensures the reliability and explainability of analytical outputs.

Beyond genomics, Virtual Lab exemplifies the application of multi-agent AI systems in experimental biomedical research. This system utilizes an AI-driven research framework, where a Principal Investigator AI leads a team of specialized agent, including a Machine Learning Specialist, Immunologist, and Computational Biologist, to design and validate nanobody binders for SARS-CoV-2 variants. The system's ability to document decision-making steps and optimize AI-driven workflows highlights the growing role of multi-agent systems in interdisciplinary research.

\begin{landscape}
\begin{table}[htbp]
\centering
\begin{adjustbox}{center, max width=\linewidth-6cm} 
\begin{tabular}{p{3cm}p{2.5cm}p{4cm}p{7cm}p{2cm}p{4cm}p{6cm}}
\toprule
Method	&	Number of agents	&	Data	&	Task and goal	&	Team structure	&	Tool use	&	Explainability	\\
\midrule
ChatNT\cite{richard2024chatnt} 	&	No agent	&	DNA, RNA, protein sequences and text data	&	Interpret biological information encoded in genome sequences and provide accurate predictions for various biological functions, such as gene expression prediction, DNA methylation, RNA stability, and protein properties.	&	NA	&	NA	&	NA	\\
DRAGON-AI\citep{toro2023dynamic}	&	No agent	&	Structured data from existing ontologies and unstructured textual data from sources like GitHub issues	&	Generate ontological terms.	&	NA	&	NA	&	NA		\\
GeneGPT\cite{jin2023genegpt}	&	No agent	&	Text data	&	 Answer genomics-related questions by directly generating API request URLs to access and retrieve relevant biomedical information.	&	NA	&	NA	&	NA		\\
Phenomics Assistant\cite{o2024phenomics}	&	Single AI agent	&	Monarch$^a$ knowledge graph	&	Enhance accessibility to complex genomic information by enabling natural language querying of the Monarch knowledge graph.	&	NA	&	Monarch Initiative API	&	The explainability of AI-generated answers by grounding them in data retrieved from the Monarch KG.	\\
VarChat\cite{de2024varchat}	&	Single AI agent	&	Scientific literature and human genomic variants	&	 Support genetic professionals by providing concise summaries of scientific literature related to specific genomic variants.	&	NA	&	Query genomic databases; find and summarize the fragmented scientific literature	&	 Informing users about the sources of its responses.		\\
ChatGSE/\newline biochatter\cite{lobentanzer2023platform}	&	Two  AI agents	&	Knowledge graph and scientific articles	&	Answer user's questions using context from knowledge graphs and scientific papers; demonstrate the usability in cell type annotation task.	&	Sequential	&	Information retrieval from knowledge graphs and the literature	&	Fact-checked and supplemented with context-specific information from documented sources.		\\
BioDiscoveryAgent\cite{roohani2024biodiscoveryagent}	&	Two  AI agents	&	Biological database (Reactome$^b$ 2022 database) and literature	&	Design genetic perturbation experiments that efficiently navigate the hypothesis space to identify a small subset of genes resulting in specific phenotypes.	&	Sequential	&	Search the biomedical literature and execute code to analyze biological datasets	&	Detailed explanations for its choices, including citing relevant literature and detailing the reasoning behind selecting specific genes for perturbation.		\\
GeneAgent\cite{wang2024geneagent}	&	Two  AI agents	&	GO, Molecular Signature Database (MSigDB$^c$), and a proteomics analysis system (NeST$^d$)	&	Generate biological process names for gene sets.	&	Sequential	&	Call web APIs that connect to biological databases	&	Providing verification reports that detail the evidence supporting or refuting each generated name.	\\
BRAD\cite{pickard2024bioinformatics}	&	Multiple AI agents	&	Online literature repositories, Enrichr$^e$ and Gene Ontology databases	&	Automate bioinformatics workflows, enhancing the speed and efficacy of tasks such as gene enrichment analysis, literature searching, and running software pipelines.	&	Hierarchical	&	Search online literature, execute code to run software pipelines, such as enrichment analyze and visualization	&	Providing context-rich answers that include references to data sources and literature.	\\
BKGAgent\cite{lin2024biokgbench}	&	Multiple AI agents	&	Clinical Knowledge and academic literature
Graph	&	The primary task is Knowledge Graph Checking, which involves querying KGs, verifying the correctness of the information using external literature or databases, and identifying factual errors.	&	Hierarchical	&	Specific tools for interacting with knowledge graphs and scientific literature	&	Agent actions and decisions are traceable and justifiable, particularly in the context of verifying scientific claims and correcting knowledge graph data.	\\
GenoAgent\cite{liu2024genotex}	&	Multiple AI agents	&	GEO$^f$ and TCGA$^g$ databases	&	Automate the analysis of gene expression data to identify disease-associated genes.	&	Hierarchical	&	Various bioinformatics tools, such as those for data normalization and statistical analysis	&	Provide explainable results by documenting the decision-making process and the steps followed in the data analysis.		\\
ProtAgents\cite{ghafarollahi2024protagents}	&	Multiple AI agents	&	Protein sequences, structural data, simulations and external databases	&	Automate and enhance the design of novel proteins with specific mechanical properties. This involves generating new proteins, analyzing their structures, and obtaining new first-principles data through physics simulations.	&	Hierarchical (dynamic, collaborative multi-agent environment)	&	Physics simulators and generative AI models, to perform tasks ranging from data retrieval to complex simulations of protein behaviors.	&	Provide explainable results by detailing the reasoning behind its decisions, the data used, and the methodologies applied.	\\
\cite{bersenev2024replicating}	&	Multiple AI agents	&	Single-cell RNA sequencing (scRNA-seq) data and literature	&	Replicate the experimental and analysis process of a scientific publication that explored gene expression relevant to SARS-CoV-2 entry into human cells. The goal is to validate the methods used in the original publication and to enhance the reproducibility and transparency of scientific research using AI.	&	Hierarchical	&	Software tools for data analysis and paper summary	&	Providing detailed breakdowns of its analytical processes and how decisions and analyses are derived, allowing for a transparent review of its methodology replication.	\\
TAIS\cite{liu2024toward}	&	Multiple AI agents	&	 TCGA, NCBI Gene and GEO databases	&	Identify disease-predictive genes from gene expression data.	&	Hierarchical	&	Various computational tools and methods integrated into the data processing and analysis workflows	&	Not detailed in the paper.	\\
Virtual Lab\cite{swanson2024virtual} & Multiple AI agents & Public protein databases and SARS-CoV-2 variant data & Design and validate nanobody binders for SARS-CoV-2 variants using AI-driven workflows. & Hierarchical & Bioinformatics tools for protein analysis & Providing explainability by structuring AI-agent meetings, documenting decisions, and presenting clear computational workflows. \\
\bottomrule
\end{tabular}
\end{adjustbox}
\caption{AI assistants in biological and biomedical domains. NA: not applicable. $^a$ \url{https://monarchinitiative.org/}. $^b$ \url{https://reactome.org/}. $^c$ \url{https://www.gsea-msigdb.org/gsea/msigdb}. $^d$ \url{https://idekerlab.ucsd.edu/nest/}. $^e$ \url{https://maayanlab.cloud/Enrichr/}. $^f$ \url{https://www.ncbi.nlm.nih.gov/geo/}. $^g$ \url{https://www.cancer.gov/ccg/research/genome-sequencing/tcga}.}
\label{tab:table1}
\end{table}
\end{landscape}

Lastly, ProtAgents showcases a multi-agent application in the design and analysis of novel proteins. By integrating real-time data from experiments and simulations, these agents can generate and analyze new proteins, adjusting their outputs based on dynamic inputs. The multi-agent system developed by Bersenev \emph{et al.} \cite{bersenev2024replicating} facilitates the replication of high-impact scientific studies by processing research papers and generating code to reproduce experiments, streamlining experimental validation and iterative scientific discovery.

Table 1 summarizes information from these studies, aligning certain elements with the HAIGoGA framework, including data, tasks, goals, AI systems and tools, and team structure (environment). The data, tasks, goals, and tools are customized for different AI assistants. In studies involving multiple agents, these agents are often organized hierarchically, with a high-level agent (e.g., planner, leader, or manager) responsible for task distribution and coordination of the analysis process.
Regarding the human-machine interface, three studies provide both GUI and CUI to facilitate human interaction with AI agents \cite{o2024phenomics, de2024varchat, lobentanzer2023platform}. The most recent work, Virtual Lab \cite{swanson2024virtual}, demonstrates the impact of human-AI collaboration through experiential evidence.
In this framework, agents can delegate tasks to other agents, as well as humans.

The cognitive abilities of LLM agents, such as perception, reasoning, and planning, enable them to potentially exhibit situational awareness. For example, the ReAct agent integrates reasoning and action, iteratively repeating this process until it determines a final response. The agent evaluates the current input along with past observations to decide the next step \cite{yao2022react}.
Additionally, several studies \cite{o2024phenomics, de2024varchat, lobentanzer2023platform, pickard2024bioinformatics} explore memory management in agent systems. These agents continuously track user interactions, dynamically recalibrating their actions based on intermediate results and feedback.

Ensuring that agent actions and predictions are traceable is crucial for explainability, as it helps build shared situational awareness and trust between humans and AI. However, a key challenge lies in balancing the amount of information provided in the final answer,making it sufficiently informative without overwhelming users with excessive details.

\subsection{A vision for the HAICoGA framework}
Through our review of current LLM agents in the biological and biomedical domains, we identified multi-agent systems as a promising approach for realizing the HAICoGA framework. Existing research primarily focuses on developing autonomous systems that minimize or even eliminate human intervention. However, such fully autonomous systems have demonstrated limited effectiveness in real-world applications \citep{liu2023agentbench, swanson2024virtual}. Therefore, it is essential to keep humans in the loop to enhance system performance and reliability \citep{m2024augmenting, swanson2024virtual}.

\begin{figure}[]
\centering
\includegraphics[width=0.9\textwidth]{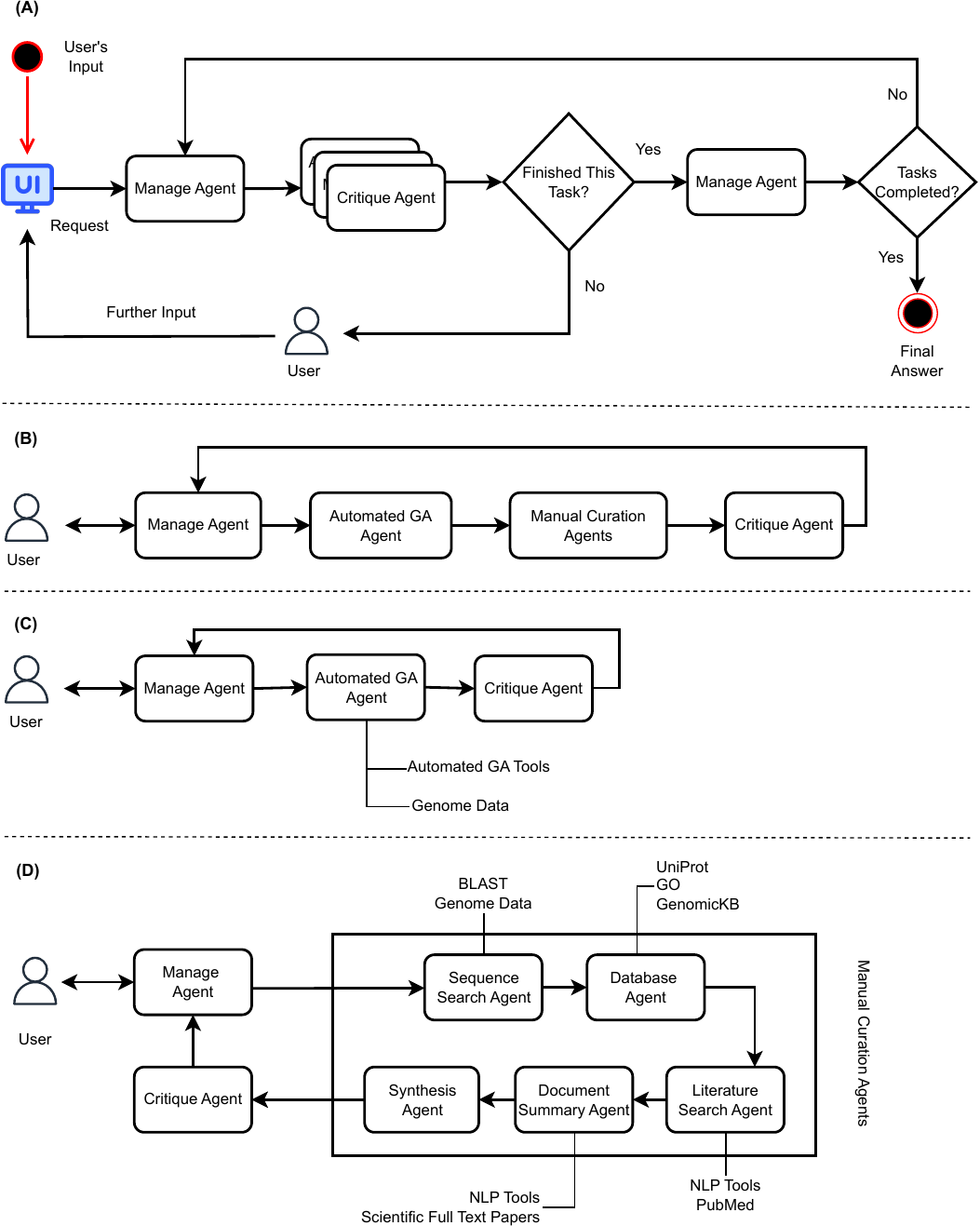}
\caption{(A) Overall multi-agent system design for human-AI collaborative genome annotation. 
Users submit a genome annotation query through an interactive user interface (UI). The UI requests the manager agent to analyze the task, decompose it into subtasks, and assign them to appropriate agents. While assisting with a subtask, an agent may request additional input from the user to complete the task successfully. The critique agent provides feedback on the outcomes, guiding the system's next steps. The manager agent monitors the global conversation history and intermediate results, updating the task plan as needed or finalizing the task and delivering the results to the user.
(B) The top synergy layer of the multi-agent system designed for HAICoGA. 
Following the practical GA workflow \cite{biology9090295}, the multi-agent system consists of a user, a manager agent, an automated GA agent, multiple manual curation agents, and a critique agent.
(C) Workflow of multi-agent collaboration in automated GA phase.
The manager agent delegates the automated GA task to the automated GA agent, which manages a customized pipeline (or an AI model) using genome data to perform specific tasks. The critique agent analyzes the results, evaluates their quality, and suggests the next steps to the manager agent. This process can be repeated iteratively until the desired outcome is achieved.
(D) Workflow of multi-agent collaboration in manual curation phase.
A manual annotation process follows the automated GA phase. Due to the complexity of manual curation, the system includes several specialized agents performing distinct roles. The sequence search agent identifies homologous genes for a target gene, for example, by running BLAST against genome sequence data. The database agent retrieves gene function annotations from various databases. The literature search agent identifies relevant scientific papers for further analysis, while the document summarization agent extracts key information from these papers. The synthesis agent compiles all relevant data and submits it to the critique agent, which reviews the information and provides suggestions, such as whether the data is sufficient to address the user's query. Finally, the manager agent either updates the task plan or generates the final response.}
\label{multi-agent}
\end{figure}

Figure \ref{multi-agent}A illustrates an example of users collaborating with a multi-agent system to annotate gene functions. Based on the user's input query, the manager agent could use a method (e.g., ReAct) for breaking down the query into subtasks and assigns them to other agents according to their capabilities (Figure \ref{multi-agent}C-D). The critique agent evaluates the quality of task results using metrics such as completeness, relevance, and other task-specific criteria, providing feedback and indicating the task's status. If necessary, agents can request additional input from the user. Once all tasks are completed, the manager agent compiles the final response and presents it to the user.

Building on the GA workflow described in the review by Ejigu \emph{et al.} \citep{biology9090295}, we propose an automated GA agent along with several agents for manual annotation (categorized as manual curation agents in Figure \ref{multi-agent}B, each assigned distinct roles, as detailed in Figure \ref{multi-agent}D). While manual annotation is often performed based on the results of automated GA, newly added manual annotations can also enhance the automated GA system by providing additional gold-standard data, enabling continuous refinement of gene annotations.

Another key strength of multi-agent systems is that it allows for the internal refinement of answers. 
In the automated GA phase (Figure \ref{multi-agent}C), the automated GA agent executes AI models and pipelines to perform specific tasks using genome data, such as predicting gene functions. The manager agent and critique agent contribute by summarizing results and providing feedback to the automated GA agent, which may prompt it to select alternative models or pipelines for gene function prediction. This iterative process enhances the quality of gene annotation. The self-improving loop continues until either the user or the manager agent decides to finalize the process and provide the final answer for the task.

The use of multiple agents also allows for specialization in the manual annotation system (Figure \ref{multi-agent}D), each assigned distinct attributes, including role, perception, and actions (tool use). These attributes enable agents to be optimized for specific domains or functions \cite{xi2023risepotentiallargelanguage}. To manually annotate an uncharacterized gene, several guidelines recommend a workflow that involves using a tool (e.g., BLAST) to identify homologous proteins, retrieving functional annotations from existing databases and recent literature, and assigning these functions to the target protein \cite{madupu2010meeting, mcdonnell2018manual, drabkin2012manual}.
Following these guidelines, the manager agent is responsible for designing this workflow and distributing tasks among specialized agents, including the sequence search agent, database agent, literature search agent, and document summary agent. The synthesis agent then aggregates the results, while the critique agent evaluates the output and provides feedback to the manager agent. Similar to the automated GA phase, the user could interrogate the results and refine prompts to continuously refine the quality of gene annotation. 

For instance, the critique agent cross-references newly generated functional annotations with existing database records, identifying potential conflicts or inconsistencies. If discrepancies arise, it can recommend further validation steps, such as retrieving additional literature sources, reevaluating sequence similarity metrics, or consulting human curators for expert review. Through this iterative validation process, the system integrates new knowledge from literature, web-lab experiments, and user inputs while aligning it with existing annotations, reducing errors propagation and improving the overall reliability of gene annotation.

Taken together, HAICoGA can be implemented as a multi-agent system supported by humans, LLM agents, AI models, and other tools working collaboratively to solve complex annotation tasks.
LLM agents, with their capabilities in skeptical learning and reasoning, are envisioned as collaborative partners that address gaps in the current GA workflow. AI models and other tools within this workflow can be executed by either humans or LLM agents. However, challenges remain in developing such a multi-agent system.

\section{Challenges in Building the HAICoGA Framework}\label{discussion}

\paragraph{Designing the architectural of a multi-agent system}
The design of LLM-based multi-agent systems requires a modular and adaptive architecture in which specialized agents collaborate dynamically through structured interaction layers. These agents, each with distinct roles, leverage LLM capabilities for reasoning and task execution while interoperating with external resources such as datasets and tools to maintain context awareness. Achieving this requires balancing autonomy and alignment, as excessive autonomy may lead to goal deviations, whereas strict alignment can hinder adaptability \citep{handler2023balancingautonomyalignmentmultidimensional}.
Furthermore, managing dependencies among agents and ensuring scalability in resource usage are critical, especially as tasks grow more complex. Mechanisms for real-time adaptation and error correction are also essential to address inconsistencies and ensure robust, goal-oriented outcomes in complex environments. Finally, challenges remain in optimizing task allocation, fostering robust reasoning through iterative debates, managing complex contextual information, and enhancing memory management \citep{he2024llmbasedmultiagentsystemssoftware}.

\paragraph{Developing novel ML/AI methods for enhancing human-AI collaboration}
LLM agents serve as collaborative partners, but recent research highlights several critical challenges that may impact their collaborative abilities.
Hallucination remains a significant issue, in which models generate plausible-sounding but factually unsupported content \citep{huang2023survey}. Since the information provided by these systems can influence decision-making, misleading outputs may propagate false beliefs or even cause harm, underscoring the need for robust mitigation strategies. Fine-tuning models for specific domains may help address domain-specific queries more effectively.
Additionally, RAG has shown promise in reducing hallucination by retrieving relevant knowledge from external sources and integrating it into the response generation process. To further enhance reliability, systems could support continuous learning, enabling dynamic updates through human feedback and evolving contexts, as exemplified by reinforcement learning from human feedback (RLHF) \citep{xi2023risepotentiallargelanguage}.

Maintaining context over extended interactions is another area where LLMs often falter, leading to incoherent responses or an inability to recall previous discussions. Vector databases offer a potential solution by enabling long-term memory management in LLM agents, allowing them to accumulate and organize memories over time. However, efficiently searching and retrieving relevant information from extensive memory stores remains challenging.
Further advancements are needed to develop mechanisms for learning and updating metadata attributes across both procedural and semantic memory types \citep{hatalis2023memory}. MemGPT \citep{packer2023memgpt} exemplifies progress in this domain by intelligently managing different memory tiers to store and retrieve information effectively during long-term conversations.

Reasoning capabilities are pivotal for LLM agents to perform complex and nuanced tasks such as problem-solving, decision-making, and planning. Explicit reasoning steps not only improve task performance but also enhance model explainability and interpretability by providing rationales for predictions.
While LLMs are primarily trained for next-token prediction, strategies like Chain of Thought (CoT) have demonstrated improvements in reasoning tasks by guiding models to articulate their reasoning explicitly. However, LLMs still face challenges in handling highly complex reasoning tasks or those involving subtle implicatures, necessitating ongoing research \citep{huang2022towards}.
 
\paragraph{Requiring multi-dimensional evaluation methods to assess the HAICoGA workflow}
Traditional GA evaluation metrics, such as coverage, precision, and accuracy, remain fundamental for assessing annotation quality \citep{ouzounis2002past, mahood2020machine}. These measures indicate better outcomes when higher values are achieved; however, they provide relative rather than absolute benchmarks due to the absence of a comprehensive genome-wide gold standard. Many annotations remain provisional, relying on computational predictions or homologous transfers from model organisms.

In HAICoGA workflows, additional dimensions, such as explainability, are crucial for evaluation. Integrating orthologous information, along with detailed protein family and domain characterizations from diverse resources, enhances the explanatory depth and reliability of annotations​ \citep{kirilenko2023integrating}. Metrics that assess explanation generation and evidence quality are essential to ensuring the transparency of AI-assisted workflows.
This aligns with frameworks for evaluating HAIC, which emphasize not only task success but also interaction quality, process dynamics, and ethical considerations \citep{fragiadakis2024evaluating}.

Furthermore, optimizing the performance of human-AI teams requires a paradigm shift from individual AI optimization to assessing team-level outcomes. Evidence suggests that the most accurate AI system does not necessarily yield the best collaborative performance​ \citep{bansal2021most}. Effective collaboration depends on dynamic task allocation, mutual learning, and trust between human and AI agents.
Metrics for evaluating such interactions must consider both qualitative factors, such as trust and satisfaction, and quantitative measures, such as decision impact and task completion time \citep{fragiadakis2024evaluating}​.

Adopting multi-dimensional evaluation frameworks, such as those emphasizing symbiotic HAIC modes, can provide holistic insights \citep{fragiadakis2024evaluating}. These frameworks should capture the dynamic, reciprocal nature of collaboration, extending beyond task success to evaluate how well humans and AI adapt to each other's strengths and limitations over time. Such comprehensive approaches are crucial for advancing the HAICoGA workflow and ensuring its alignment with both scientific rigor and practical utility​.

\paragraph{Designing intuitive and interactive interfaces to facilitate human-AI collaboration} 
To investigate the challenges and opportunities in CUIs, we developed a chatbot prototype for curating information in gene functional annotation. Additionally, we proposed applying conjoint analysis, a behavioral science method, to quantify the relative importance of four design features that influence users' trust in the system \cite{McGrath2025}.

Initial testing of the prototype suggests that LLM agents have the potential to serve as valuable tools for collaborative genome annotation (GA) when combined with human expertise. However, further research is needed to enhance their trustworthiness, particularly by improving explainability and providing confidence measures for AI-generated predictions \cite{McGrath2025}.

To support these capabilities, future work will focus on integrating a dedicated graphical user interface (GUI) with the chatbot, particularly for structural annotation. Users will be able to interact with a genomic viewer using gestures or cursor movements to select, drag, or draw annotations. Developing the right interface will be best served by taking a participatory or user-centred design approach and incorporating input from GA experts from the outset. 

\section{Conclusion}\label{conclusion}
In this paper, we first analyzed the pros and cons of automated GA methods and manual curation tools. We found that while automated GA methods generate annotations quickly, they have limitations, such as inaccurate gene predictions. On the other hand, manual curation can be highly accurate but requires intensive human labor and time.
A human-AI collaborative genome annotation approach is necessary to leverage the strengths of both humans and AI, leading to more accurate and efficient GA.

Bringing together prior work in automated GA and manual curation, we then proposed the conceptual framework of HAICoGA. Our work bridges the gap between GA and human-AI collaborative communities, envisioning new possibilities in this multidisciplinary field.
The emergence of LLM agents presents significant opportunities to realize HAICoGA workflows. However, many challenges and open questions remain in LLM agent research.
The HAICoGA framework is still in its early stages of development, but it represents a step toward a comprehensive and efficient human-AI collaborative workflow for real-world applications in the future.

\section{Glossary}
\emph{Genome annotation (GA)} is the process of identifying and characterizing functional elements within a genome, including genes, regulatory regions, and other biologically significant sequences. It involves the use of computational methods, such as machine learning (ML) and heuristic-based approaches, as well as manual curation by experts to improve accuracy. GA is essential for understanding gene functions, predicting protein structures, and exploring evolutionary relationships across species.

\emph{Artificial Intelligence (AI)} refers to the simulation of human intelligence in machines, enabling them to perform tasks such as reasoning, learning, problem-solving, and decision-making. AI encompasses various techniques, including ML, deep learning, and natural language processing (NLP), to analyze complex data and automate decision-making. In GA, AI is used to enhance the efficiency of gene prediction, functional annotation, and data integration by processing large-scale biological datasets with minimal human intervention.

\emph{Machine Learning (ML)} is a subset of AI that enables computers to learn patterns from data and make predictions or decisions without being explicitly programmed. In GA, ML algorithms are used to classify genes, predict functional elements, and enhance annotation accuracy by analyzing large-scale genomic datasets. ML approaches include supervised, unsupervised, and reinforcement learning, leveraging statistical models and neural networks to improve biological data interpretation.

\emph{Manual curation}, also known as manual annotation, refers to the process in which human experts review, refine, and validate genome annotations to ensure accuracy and biological relevance. This process involves analyzing computationally generated annotations, resolving ambiguities, and incorporating insights from experimental data and scientific literature. 

\emph{Human-AI collaboration (HAIC)} refers to the dynamic interaction between humans and AI systems, where both work together toward shared objectives by leveraging their complementary strengths. Unlike automation, where AI operates independently, or augmentation, where AI enhances human capabilities, HAIC involves a continuous exchange of information, decision-making, and adaptation over time.

\emph{Knowledge graphs (KGs)}
are structured representations of relationships between biological entities, such as variants, genes, proteins, pathways, phenotypes, and diseases. They encode known interactions and associations in a graph format, where nodes represent entities and edges denote relationships. KGs facilitate data integration, reasoning, and discovery in genomics by linking heterogeneous biological information sources.

\emph{Large language models (LLMs)} are AI models trained on massive datasets of text and code. They can generate human-quality text, translate languages, understand user instructions for task procedures \citep{wen2023empowering, singh2023progprompt}, use external tools \citep{schick2023toolformer}, and answer user questions based on specific contexts \citep{brown2020language}.

\emph{AI agent} is an autonomous or semi-autonomous entity within a multi-agent system that performs specific tasks, interacts with other agents, and operates based on predefined rules, learned behaviors, or external inputs.  Agents may specialize in different roles, such as task management, data retrieval, reasoning, or quality assessment, and they communicate within structured frameworks to enhance human-AI collaboration.

\section*{Competing interests}
No competing interest is declared.

\section*{Key points}\label{key-points}

\begin{itemize}
\item
  While genome annotation is complex and challenging, heavy reliance on
  automated methods can introduce errors.
\item
  Manual curation is necessary for accurate annotations but requires
  significant time and effort.
\item
  Our novel contribution is HAICoGA, the first conceptual framework for human-AI collaborative genome annotation.
\item
  We further present a example of HAICoGA framework and future research directions in realize this framework.
\end{itemize}

\bibliographystyle{unsrt}
\bibliography{bibliography.bib}


\begin{biography}{}{\author{Xiaomei Li} is a postdoctoral fellow at CSIRO, specializing in
		the development of frameworks and methodologies that enhance human-AI
		collaborative intelligence within the domain of Genomics and
		Bioinformatics.}
\end{biography}

\begin{biography}{}{\author{Alex Whan} is a research scientist at CSIRO, where he focuses on
the development of approaches to store and integrate experimental data
to improve analysis and insight from biological systems.}
\end{biography}

\begin{biography}{}{\author{Meredith McNeil} is a team leader at CSIRO, where her focus is
on applying advances in genomics, phenomics, and computational biology
to assist in the development of improved crop varieties for Australian
agriculture.}
\end{biography}

\begin{biography}{}{\author{David Starns} is a genome curation and annotation scientist for
the VEuPathDB database resource based at the University of Liverpool.
David focuses on parasite, fungal and vector Eukaryote genomes and
facilitates scientific community engagement.}
\end{biography}

\begin{biography}{}{\author{Jessica Irons} is a behavioural scientist in CSIRO's Data61,
researching the development of collaborative human-AI workflows, drawing
on principles from human factors and cognitive psychology.}
\end{biography}

\begin{biography}{}{\author{Samuel C. Andrew} is a research scientist at CSIRO,
    exploring how transcriptomic responses to stress and functional traits can be used to understand the thermal tolerance capacity of crops and wild plants.}
\end{biography}

\begin{biography}{}{\author{Rad Suchecki} spearheaded the efforts to develop workflows for human-AI collaborative discovery in genomics and bioinformatics during his time as a research scientist and the Genomics and AI team leader at CSIRO Ag \& Food. 
He has since joined Alkahest Inc., as a Sr Computational Biologist working with healthcare and omics data.}
\end{biography}

\end{document}